\documentclass[%
journal=jpccck,
manuscript=article,
amsmath,
amssymb
]{achemso}

\usepackage[version=3]{mhchem} %

\usepackage{graphicx}
\usepackage{dcolumn}
\usepackage{bm}
\usepackage{natbib}
\usepackage{mciteplus}
\usepackage{hyperref}
\usepackage{color}
\newcommand{\dfdx}[2]{\left(\frac{\partial #1}{\partial #2}\right)}
\newcommand{\avdbl}[1]{\left\langle\left\langle{#1}\right\rangle\right\rangle}

\author{Xipeng Wang}
\affiliation[Chinese Academy of Sciences]{Institute of Physics, Chinese Academy of Sciences,  Beijing 100190, China}
\alsoaffiliation{Department of Chemistry, University of Cambridge, Cambridge, UK}
\author{Sim\'on Ram\'irez-Hinestrosa}
\affiliation[University of Cambridge]{Department of Chemistry, University of Cambridge, Cambridge, UK}
\author{Daan Frenkel}
\affiliation[University of Cambridge]{Department of Chemistry, University of Cambridge, Cambridge, UK}
\email{df246@cam.ac.uk}

\title{Using Molecular Simulation to Compute Transport Coefficients of Molecular Gases} 

\begin{document}


\date{\today}

\begin{abstract}
 The existing kinetic theory of gases is based on an analytical approach that becomes intractable for all but the simplest molecules. 

Here we propose a simple numerical scheme to compute the transport properties of molecular gases in the limit of infinite dilution. 
The approach that we propose is approximate, but our results for the diffusivity $D$, the viscosity $\eta$ and the thermal conductivity $\lambda$ of hard spheres, Lennard-Jones particles and rough hard spheres, agree well with the standard (lowest order) Chapman-Enskog results.  
We also present results for a Lennard-Jones-dimer model for nitrogen, for which no analytical results are available.
In the case of poly-atomic molecules (we consider n-octane), our method remains simple and gives good predictions for the  diffusivity and the viscosity. Computing the thermal conductivity of poly-atomic molecules requires an approximate treatment of their quantized internal modes.   
We show that a well-known approximation that relates $\lambda$ to $D$ and $\eta$, yields good results.    
We note that our approach should yield a lower limit to the exact value of $D$, $\eta$ and $\lambda$. 
Interestingly, the most sophisticated (higher-order) Chapman-Enskog results for rough hard spheres seem to violate this bound.
\end{abstract}

\maketitle


\section{Introduction}\label{sec:Intro}
The theoretical framework for computing the transport coefficients of gases was developed by Boltzmann in 1872~\cite{bol721} . 
General methods to compute the transport coefficients for spherical particles  were developed by Chapman and Enskog, just over a century ago (see e.g. ref.~\citenum{cha701}). 
At that time (almost half a century before the advent of computer simulations) the focus was on obtaining analytical expressions for certain classes of pair potentials. 
However, the more general case of molecules with internal degrees of freedom cannot be solved with such an  analytical approach, except in a small number of very simple, or highly simplified cases. 
Strangely, this state of affairs has barely changed during the past half century.  
One reason may be that the Chapman-Enskog formalism is not for the faint-hearted:
as Chapman himself (quoted in the Observer of July 7, 1957, p11), said about his book~\cite{cha701}: {\em It is very-heavy going and  ``like chewing glass''}.\\
The apparent lack of simulation-based studies of molecular transport properties is all the more surprising as there are literally thousands of papers on molecular trajectory calculations (a review from 1974 already lists 245 references~\cite{por741}),  many of them aimed at studying molecular scattering and collision-induced relaxation of intra-molecular degrees of freedom. 
Yet, most of these thousands of papers do not aim to compute transport properties. 
Notable exceptions are the papers by Nyeland et al.~\cite{nye841} on the transport properties of gaseous nitrogen, and the paper by Viehland et al.~\cite{vie961} on the mobility of $NO^+$ in Helium. 
Whilst there are a few other papers reporting similar calculations, we have been unable to find systematic molecular-trajectory calculations of the transport properties of molecular gases containing more than 2 atoms. 
The lack of such papers is not surprising, as the calculations are far from simple (as mentioned explicitly in ref.~\citenum{vie961}).

Surprisingly, the situation is better for liquids than for gases because the advent of Molecular Dynamics simulations made it possible to compute a variety of transport coefficients as integrals of time correlation functions, using the so-called Green-Kubo relations (see e.g.~\citenum{han131}). 
However, this did not help the problem of transport in gases at low densities, as the relevant time-correlation functions decay increasingly slowly as the density is decreased, and as a consequence the statistical error in the transport coefficients diverges as the density tends to zero~\cite{zwa691} . 
As a consequence, the number of Molecular Dynamics calculations of transport coefficients of dilute gases is very small.  
An MD study by Lee and Kim of the the transport coefficients of simple diatomic gases at atmospheric pressure~\cite{lee141} provides an illustration of the technical challenges of a ``brute force'' MD approach: these simulations are reasonably long ($\mathcal{O}$(10) ns), yet 10 ns  corresponds to only $\mathcal{O}$(50) correlation times. 
Hence, the statistical error in the transport coefficients (except the diffusivity) is expected to be appreciable (although a non-standard analysis may mitigate the problems somewhat~\cite{kra111}).

Summarizing this discussion:  there seems to be no efficient, generally applicable method to compute the transport coefficients of dilute poly-atomic, molecular gases for which Chapman-Enskog-style analytical approaches become intractable. 
This is a real problem, as knowledge of the transport coefficients of dilute gases is important in many fields, such as atmospheric science. 
The aim of the present paper is to fill this gap. 
We consider the calculation of the most important transport coefficients of a pure gas of molecules with finite ranged, but otherwise arbitrary, intermolecular potentials. 
As we show, these expressions lend themselves to  simple numerical simulations.  
\subsection{Green-Kubo relations}
Rather than starting from the Boltzmann equation, we start from the Green-Kubo  (GK) expressions (see e.g.~\citenum{han131}) for the most important transport coefficients: the diffusivity $D$, the shear viscosity $\eta$ and the thermal conductivity $\lambda$.
The GK relations relate the various transport to an integral of the time auto-correlation function of the appropriate flux, as listed below.
\begin{equation}\label{eq:mdgreenkuboD}
D=\frac{1}{3}\int_0^\infty d\;  \tau \; \left \langle {\bf v}(\tau)\cdot {\bf v}(0)\right \rangle = \int_0^\infty d\;  \tau \; \left \langle v^x(\tau) v^x(0)\right\rangle
\end{equation}
The Green-Kubo relation for the  shear viscosity $\eta$ is:
\begin{equation}\label{eq:mdeta}
\eta=\frac{1}{Vk_BT}\int_0^\infty d\;  t\; \left \langle \sigma^{xy}(0)\sigma^{xy}(t)\right \rangle 
\end{equation}
with
\begin{equation}\label{eq:mdsigmaxy}
\sigma^{xy}=\sum_{i=1}^N \left(m_i v_i^xv_i^y+
\frac{1}{2}\sum_{j\ne i}x_{ij} f_y(r_{ij})\right);
\end{equation}
Finally, the thermal conductivity $\lambda$, can be obtained from
\begin{equation}\label{eq:mdlambda}
\lambda=\frac{1}{Vk_BT^2}\int_0^\infty d\;  t\; \left \langle j^Q_z(0)j^Q_z(t)\right \rangle 
\end{equation}
where the heat flux $j^Q_z$ ia given by
\begin{equation}\label{eq:je}
j^Q_z=
\frac{d\;  }{dt}\sum_{i=1}^N z_i\left[\left(\frac{1}{2}m_i v_i^2 +E_{\rm int}^{(i)}-h_i\right)+
\sum_{j\ne i} v(r_{ij})\right]\;,
\end{equation}
where $E_{\rm int}^{(i)}$ is the internal (vibration, rotation, electronic) energy of molecule $i$ and  $h_i$ is its the average enthalpy ($h_i$ = $\frac{5}{2}k_BT + <E_{\rm int}^{(i)}>$).
These expressions are generally valid for pairwise additive potentials, and are used extensively in Molecular Dynamics simulations to compute the transport coefficients of dense fluids. 
However, in the dilute gas limit, the GK integrals -- although still correct -- converge slowly, and the statistics become poor. 
The problem is analogous to computing the second virial coefficient $B_2$ in simulation: yes, it would be possible to compute $B_2$ by evaluation the compressibility factor of a dilute gas, but the method does not work well, and direct computation of the explicit expression for $B_2$ is preferable.
The same holds for transport coefficients of dilute gases. 
\subsection{Green-Kubo relations for dilute gases}
In dilute gases, the GK expressions for $\eta$ and $\lambda$ simplify, because we can ignore the contribution to the fluxes due to terms involving the intermolecular interactions. 
That is:
\begin{equation}\label{eq:sigmaxy_dilute}
\sigma^{xy}=\sum_{i=1}^N m_i v_i^xv_i^y
\end{equation}
and
\begin{equation}\label{eq:je_dilute}
j^Q_z= \sum_{i=1}^N \hat{\bf z}.{\bf v}_i\left[\frac{1}{2}m_i v_i^2+E_{\rm int}^{(i)}-h_i\right]
\end{equation}
The expression for the diffusion coefficient remains unchanged. It is worth pointing out that, when using the Green-Kubo expression for one-component fluids, the constant $h_i$ is usually omitted. 
However, this is only allowed if we can work in a reference frame where $\sum_i {\bf v}_i$=0. But later on, when we compute the heat flux correlation function by considering independent binary collisions, there is no such cancellation on a term-by-term basis: hence the $h_i$ should be retained. 

The next simplification comes from the fact that, in very dilute gases, the ``molecular chaos'' assumption holds: molecules undergo infrequent collision events with a typical duration $\tau_c$ separated by time intervals between collisions of  order $\tau_{bc}$, such that $\tau_{bc}\gg \tau_c$. 
For molecules with attractive interactions, a small fraction of all collisions (``orbiting collisions'') may take arbitrarily long, making the time-scale separation between $\tau_{bc}$ and $\tau_c$ questionable. 
The simulation time limits the largest values of $\tau_c$ that can be probed. 
Note that we can ignore true  bound states that can only be broken up by a collision with a third particle, as the concentration of bound pairs vanishes quadratically at low densities. 

The molecular chaos assumption implies that successive collision events are uncorrelated. 
In the low-density limit, this hypothesis is justified (at higher densities  mode-coupling effects become important and the molecular chaos approximation breaks down).  
\subsection{BGK approximation}
We now make an important additional assumption, namely that the relevant correlation functions decay as single exponentials. 
This is {\em not} correct, not even at infinite dilution. In the context of the solution of the Boltzmann equation, the ``single-exponential'' approximation is usually referred to as the  BGK (Bhatnagar, Gross and Krook~\cite{bha541}) approximation. 
Strictly speaking, we do not have to use a BGK-like approximation, but if we do not, the present scheme loses much of its simplicity.
In what follows, we shall refer to the ``single-exponential'' approximation, as the BGK approximation.

One more comment is in order: 
The range of interaction between molecules is typically not finite, hence strictly-speaking, the duration of a collision is not finite. 
Whilst this is true, we should expect that in many cases of practical importance, the transport properties of particles with a finite interaction range $r_c$ will approach those of particles with the full interaction, provided we choose $r_c$ large enough. 
In what follows, we will therefore always assume that the inter-particle interactions have a finite range $r_c$. 
We stress that such an approach will not work for molecules interacting through long-ranged  interactions. 
However, extension of our numerical method to systems with true long-ranged interactions between uncharged molecules, is straightforward. 
Such an extension is achieved by a change of variables in the collision integrals (see SI).

Let us next consider the three correlation functions mentioned above. 
We denote these correlation functions by $C_\alpha(t)$, where $\alpha$ may refer to the velocity, stress or heat flux. 
It should be emphasised that  the velocity is a singe-particle property and hence it is only necessary to follow the effect of a collision on one of the two collision partners. 
However, stress and heat flux are collective properties, hence we have to compute how a collision changes the stress or heat flux of a {\em pair} of particles. 
In the regime where molecular chaos holds, we can write (within the BGK approximation):
\begin{equation}
C_\alpha(t)=<J_\alpha^2>e^{-t/\tau_\alpha}\;,
\end{equation}
where $J_\alpha$ denotes the flux associated with $\alpha$, and hence
\begin{equation}
\int_0^\infty dt\; C_\alpha(t)=<J_\alpha^2>\tau_\alpha 
\end{equation}
Our aim is therefore to compute $\tau$ (for dilute gases the equilibrium average $<J_\alpha^2>$ can be evaluated analytically -- or almost analytically, in the case of heat fluxes of particles with internal degrees of freedom).
To obtain an expression for $\tau$, we note that
\begin{equation}
\dfdx{C_\alpha(t)}{t}_{t=0+}=-\frac{<J_\alpha^2>}{\tau_\alpha}
\end{equation}
Hence,
\begin{equation}
\int_0^\infty dt\; C_\alpha(t)=-<J_\alpha^2> \left(\frac{<J_\alpha^2>}{\dot{C}_\alpha(0+)}\right)
\end{equation}
\begin{figure*}[htb]
\centering
\includegraphics[width=\linewidth]{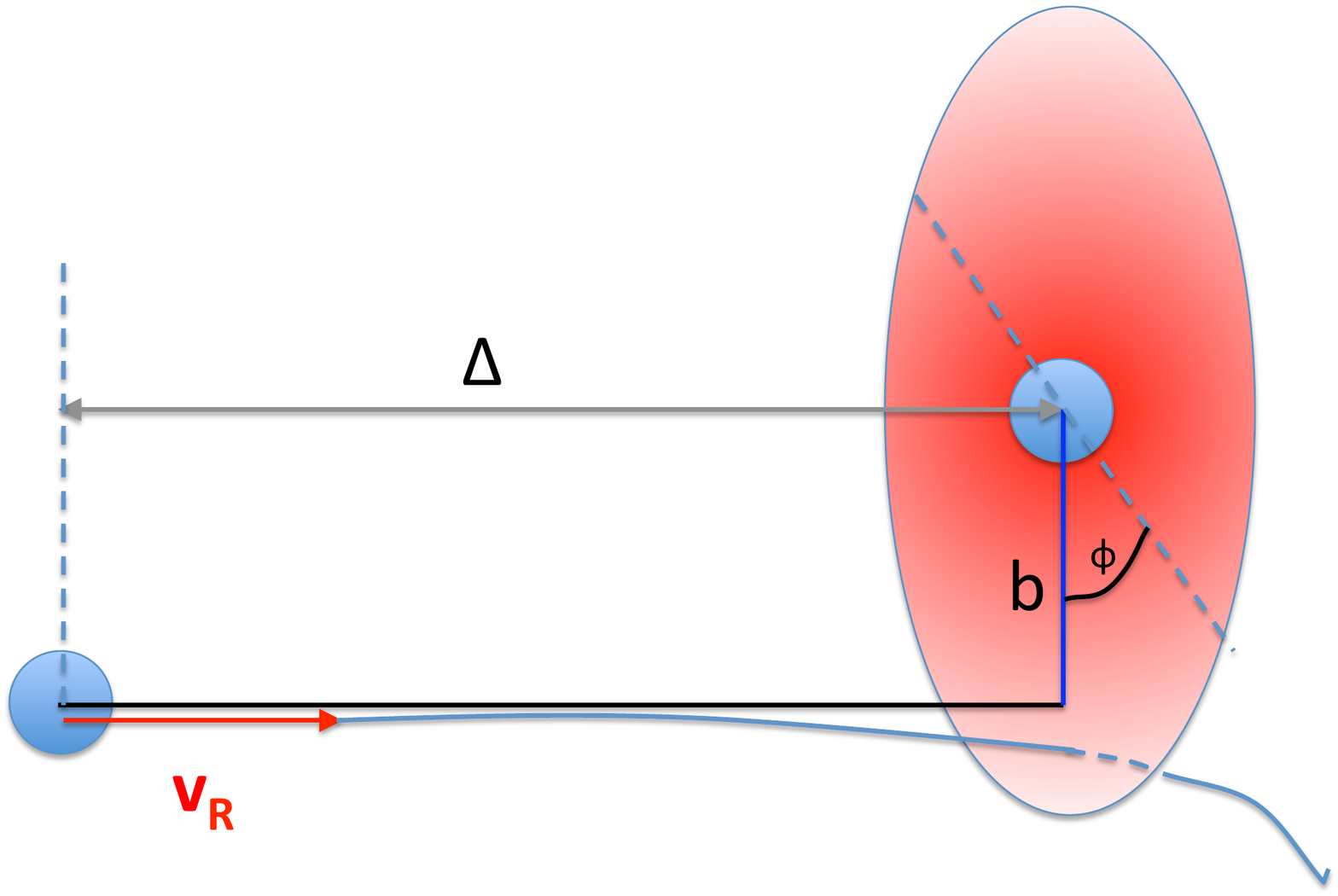}
\caption{Sketch of the collision geometry of two particle with relative velocity $v_{\rm rel}$  and impact parameter $b$. 
If the particles are not spherically symmetric, we should average over all particle orientations: in that case, the angle $\phi$ is unimportant  and can be taken equal to 0. }
\label{fig:collision}
\end{figure*}
Note that the decay of $C_\alpha(t)$ is due to the  (uncorrelated) collision events.
We can therefore write $\dot{C}_\alpha(0+)$ as
\begin{equation}
\dot{C}_\alpha(0+) = < J_\alpha \Delta_c J_\alpha>_c\times \Gamma_c
\end{equation}
where $\Delta_c J_\alpha$ denotes the change in $J_\alpha$ in a single collision, $<\cdots>_c$ denotes averaging over all collision conditions (see below), and $\Gamma_c$ is the collision frequency. 
The number of collisions experienced per unit time by a single particle is equal to $\Gamma^{(1)}$=$\rho\pi r_c^2 <v_{\rm rel}>$, where $\rho$ is the number density and $r_c$ is the maximum range of the interaction between collision partners~\footnote{In fact, the collision cross section may be chosen larger than $\pi r_c^2$, because this factor drops out of the overall expression. 
However,  a smaller collision cross section than $\pi r_c^2$ is not allowed.}, and $v_{\rm rel}$ is the relative velocity of two collision partners. 
The total collision frequency of $N$ particles is $\Gamma^{(N)} = (N/2) \Gamma^{(1)}$. 
The factor 1/2 is needed because every collision involves {\em two} particles.

In the case of diffusion, $J_\alpha$ is the velocity of a single particle in (say) the $x$-direction, and hence
\begin{equation}
\dot{C}_D(0+) = < v_x \Delta_c v_x>_c\times \Gamma^{(1)}\;.
\end{equation}
However, in the case of viscosity and heat conductivity, $J_\alpha$ is an $N$-particle current.
Yet, when we consider $< J_\alpha \Delta_c J_\alpha>_c$, only the terms in $J_\alpha$ due to the collision partners are correlated with $\Delta_c J_\alpha$, which involves those particles.  
In what follows, we will  denote that part of the current that is changed in a collision by $j_\alpha$.

Let us consider the variation of $C_\alpha(t)$ in the interval $0\le t\le t^*$, where $t^*$ is a time interval much larger than the duration of a collision, but much shorter than the average time between collisions experienced by a typical particle. 
We define $\Delta c_\alpha(t^*)$, the average change of $J_\alpha(0)J_\alpha(t)$ due to {\em a single} collision event. 
As mentioned above, a single collision will only affect the flux due to one (diffusion) or two ($\eta$ or $\lambda$) particles. 
Using our definition for  $j_\alpha$, we can then write:
\begin{equation}
\Delta (j_\alpha(0)j_\alpha(t^*))= 
 j_\alpha(0)\left(j_\alpha(t^*)-j_\alpha(0)\right) =  j_\alpha(0)\Delta j_\alpha(t^*)
\end{equation}
where the last equality defines $\Delta j_\alpha(t^*)$
\begin{equation}
\Delta j_\alpha(t^*)\equiv j_\alpha(t^*)-j_\alpha(0)\;.
\end{equation}
To obtain  $\dot{C}_\alpha$, we have to consider all collision conditions (relative speed, impact parameter, internal energy) and take and average. 

The rate of change of $\left\langle j_\alpha(0)\Delta j_\alpha(t^*)\right\rangle$ is equal to the average change in $j_\alpha(0)j_\alpha(t^*)$ during an interval of length $t^*$.  
This can be written as the average change in $j_\alpha(0)j_\alpha(t^*)$ during a collision,  multiplied by the number of collisions per time interval $t^*$.
We define a collision as any event where two particles that were initially not interacting cross through each other's interaction zone. 
For particles with a relative velocity $v_{\rm rel}$, the number of such events, in an interval $t^*$ is $\rho t^*v_{\rm rel} \pi r_c^2$ (i.e. the number of particles in a cylinder with cross-section $\pi r_c^2$ and height $t^* v_{\rm rel}$ ). 
Note that it is convenient to use a coordinate system where one axis (say $X$) is along the direction of the relative velocity of the collision pair. 
It is important to distinguish between the lab-based coordinate frame $x,y,z$ and the collision frame $X,Y,Z$. 
Later we will have to average over all orientations of $X,Y,Z$ with respect to the lab frame. 
The distribution of the magnitude of the relative velocity is 
\begin{equation}
P(v_{\rm rel}) = 4\pi\left(\frac{\beta\mu}{2\pi}\right)^{3/2} v^2_{\rm rel} \exp(-\frac{1}{2}\beta\mu v^2_{\rm rel}) \;,
\end{equation}
where $\mu$ $\equiv$ $ m_1m_2/(m_1+m_2)$ is the reduced mass of particles with masses $m_1$ and $m_2$.
In this coordinate frame (see Fig.~\ref{fig:collision}), we initially position the two collision partners at a distance $\Delta$ in the direction of ${\bf v}_{\rm rel}$, whilst the distance in the perpendicular direction is given by the (2D) polar coordinates $b$ and $\phi$.
The value of $\Delta j_\alpha(t^*)$ depends both on the value of $v_{\rm rel}$ and on the ``impact parameter'' $b$, i.e. the magnitude of the projection of the initial inter-particle distance of the plane perpendicular to $v_{\rm rel}$.

Our aim is to compute the rate of change of $\left\langle j_\alpha(0)\Delta j_\alpha(t^*)\right\rangle$ due to single collision events. 
The averaging is done over all impact parameters and over all relative velocities (and, in the heat flux case, over all center-of-mass speeds):
\begin{eqnarray}\label{eq:dotc}
 \dot{c}_\alpha&=& \rho \left\langle j_\alpha(0)\Delta j_{\alpha;c}\right\rangle_1\nonumber\\ 
 &=&\rho \int dX_{\rm int}P(X_{\rm int})
 \int_0^{r_c}\pi db^2\int_0^\infty dv_{\rm rel} \;  v_{\rm rel}  P(v_{\rm rel})
 j_\alpha(0)\Delta j_{\alpha;c}(b, v_{\rm rel};X_{\rm int}) \;,
\end{eqnarray}
where $\Delta j_{\alpha;c}(b, v_{\rm rel};X_{\rm int}) $ denotes the change in  $j_\alpha$ during one collision with impact parameter $b$ and relative velocity $v_{\rm rel}$, given that the initial state of the collision pair is given by $X_{\rm int}$: 
$X_{\rm int}$ is used to denote  all the internal degrees of freedom of both particles {\em before} the collision, {\em and} the center-of-mass velocity of the collision pair (as we shall see below, this quantity is only important for the thermal conductivity). 

Note that 
\begin{equation}
\Gamma^{(1)}=\rho\pi r_c^2<v_{\rm rel}> = \rho\int dX_{\rm int}P(X_{\rm int})
 \int_0^{r_c}\pi db^2\int_0^\infty dv_{\rm rel} \;  v_{\rm rel}  P(v_{\rm rel})
\end{equation}
We can therefore rewrite Eqn.~\ref{eq:dotc} as
\begin{eqnarray}\label{eq:dotc3}
 \dot{c}_\alpha &=& \Gamma^{(1)} \frac{\int dX_{\rm int}P(X_{\rm int} )
 \int_0^{r_c}\pi db^2\int_0^\infty dv_{\rm rel} \;  v_{\rm rel}  P(v_{\rm rel})
 j_\alpha(0)\Delta j_{\alpha;c}(b, v_{\rm rel};X_{\rm int})}{<v_{\rm rel}> \pi r_c^2}\nonumber\\
 &\equiv& \Gamma^{(1)} \langle j_\alpha(0)\Delta j_{\alpha}\rangle_c
 \;.
\end{eqnarray}
Eqn.~\ref{eq:dotc3} is convenient because it expresses $\dot{c}_\alpha$ as the product of a collision frequency and an average contribution per collision. 

For convenience,  we have assumed that the interaction potential has a finite range $r_c$, such that we can limit the integration over impact parameters to values less than $r_c$.

It is important to note that $j_\alpha(0)$ refers to the state {\em before} the collision particles interact, and $\Delta j_{\alpha;c}$ denotes the difference between $j_\alpha(0)$ and the value of $j_\alpha$ {\em after} the collision. 
Hence, neither  $j_\alpha(0)$ nor $\Delta j_{\alpha;c}$ depend explicitly on time.  
More importantly, $j_\alpha(0)$ and $\Delta j_{\alpha;c}$ contain only kinetic contributions ({\em e.g.} $v^x$ or $v^xv^y$) and no inter-molecular interaction terms. 

The next step is to separate the particle velocities in a center-of-mass part and the relative velocity: 
\begin{equation}
{\bf v}^{(1)} = {\bf v}_{\rm CM} + \frac{m_2}{m_1+m_2} {\bf v}_{\rm rel}\;,
\end{equation}
and
\begin{equation}
{\bf v}^{(2)} = {\bf v}_{\rm CM} - \frac{m_1}{m_1+m_2} {\bf v}_{\rm rel}\;,
\end{equation}
 where 
 \begin{equation}
 {\bf v}_{\rm CM}\equiv \frac{m_1{\bf v}^{(1)}_{\rm rel}+m_2{\bf v}^{(2)}_{\rm rel}}{m_1+m_2}\;,
 \end{equation}
 and 
 \begin{equation}
  {\bf v}_{\rm rel}\equiv {\bf v}^{(1)}_{\rm rel}-{\bf v}^{(2)}_{\rm rel} \;.
 \end{equation}

We can always decompose $j_\alpha$ in $j_\alpha^{\rm CM}$ and $j_\alpha^{\rm rel}$. Importantly,  $j_\alpha^{\rm CM}$ is unchanged in a collision, and $\Delta j_{\alpha;c}$ is independent of $j_\alpha^{\rm CM}(0)$~\footnote{As we shall when discussing the heat flux, it is a bit more subtle: $\Delta j_{\alpha;c}$ may still depend on the magnitude of $ {\bf v}_{\rm CM}$, but not on its orientation.}.
Hence
\begin{equation}\label{eq:dotc2}
\dot{c}_\alpha =\int dX_{\rm int} P(X_{\rm int})\int_0^{r_c}\pi db^2\int d{\bf v}_{\rm CM}\; P({\bf v}_{\rm CM})
\int_0^\infty dv_{\rm rel} \; (\rho v_{\rm rel})  P(v_{\rm rel})
 j_\alpha(0)\Delta j_{\alpha_c}\;,
\end{equation}
The center-of-mass velocity still shows up in this equation, but this dependence is only needed for the heat conductivity.
In what follows, we will assume for simplicity that we are considering a pure substance. 
In that case $m_1=m_2\equiv m$.
Let us now consider the expressions for $D$, $\eta$ and $\lambda$ using $x,y,z$ to denote directions in the lab frame.
In the case of diffusion, we have 
\begin{equation}
j_D=v^x 
\end{equation} 
For $\eta$ we have 
\begin{equation}
j_\eta=m \sum_{i=1}^2 v_i^xv_i^y
\end{equation}
and for $\lambda$
\begin{equation}
j_\lambda= \sum_{i=1}^2  v_i^x \left(\frac{1}{2}mv_i^2+ E_{\rm int}^{(i)}-h_i\right) \;,
\end{equation}
where $E_{\rm int}^{(i)}$ denotes the total (kinetic plus potential) internal energy of molecule $i$. 
The internal energy comprises rotational and vibrational (and possibly electronic) contributions. 
Note that for $\eta$ and $\lambda$, the corresponding expression for $\alpha$ contains the contribution of both collision partners. 

For $\dot{C}_D$, we can now write: 
\begin{equation}\label{eq:dotcD}
\dot{c}_D =\frac{\rho\Gamma^{(1)}}{4}\langle v_{\rm rel}^x \Delta v_{\rm rel}^x\rangle_c \;.
\end{equation}
The factor (1/4) in front of Eqn.~\ref{eq:dotcD} is due the the fact the velocity of a {\em single} particle is ${\bf v}_{\rm CM}\pm (1/2){\bf v}_{\rm rel}$. 
$\dot{c}_D$ does not depend on ${\bf v}_{\rm CM}$, but the factor $\pm(1/2)$ in front of ${\bf v}_{\rm rel}$ enters quadratically in Eqn.~\ref{eq:dotcD}.
When we average the orientations of the collision frame with respect to the lab frame, we can write 
\begin{equation}\label{eq:orient_avg}
v_{\rm rel}^x\left(
\Delta  v_{\rm rel}^x\right)=
\frac{1}{3}
v_{\rm rel}(0)\left[v_{\rm rel}(1)\cos\theta - v_{\rm rel}(0)\right]\;,
\end{equation}
where we have averaged over all orientations of $v_x$ in the lab frame. 
We have defined
\begin{equation}
\cos\theta\equiv \hat{{\bf v}}_{\rm rel}(0)\cdot\hat{{\bf v}}_{\rm rel}(1)\;,
\end{equation}
and $v_{\rm rel}(0)$, $v_{\rm rel}(1)$ denote the relative velocities before and after the collision.
In eqn.~\ref{eq:dotcD}, the dependence on ${\bf v}_{\rm CM}$ has disappeared because during a collision $\Delta {\bf v}_{\rm CM}$ = 0 and, in addition, ${\bf v}_{\rm CM}(0)$ is not correlated with $ \Delta {\bf v}_{\rm rel}$.
Note that we have assumed that ${\bf v}_{\rm rel}$ is initially in the $+x$ direction.  
Although the internal degrees of freedom do not enter explicitly in the expression for $\dot{c}_D$, they enter implicitly, because they may affect the post-collisional relative velocity. 
This is different for particles with no internal degrees of freedom: in that case all collisions are elastic, and the magnitude of the relative velocity is not changed by the collision.
From Eqns.~\ref{eq:dotcD} and \ref{eq:orient_avg} it follows that
\begin{equation}\label{eq:dotcDA}
\dot{c}_D =\frac{\Gamma^{(1)}}{12}\langle
v_{\rm rel}(0)\left[v_{\rm rel}(1)\cos\theta - v_{\rm rel}(0)\right]
\rangle_c \;,
\end{equation}
For $\eta$ we have 
\begin{equation}\label{eq:dotcEta}
\dot{c}_\eta =\frac{\rho m^2}{4}
\langle v_{\rm rel}^xv_{\rm rel}^y\left(
\Delta  v_{\rm rel}^xv_{\rm rel}^y\right)\rangle_c\;,
\end{equation}
For the viscosity $\eta$ and the thermal conductivity $\lambda$, the distinction between $\dot{C}_\alpha$ and $\dot{c}_\alpha$ is important:  $\dot{C}_\eta = (N/2) \dot{c}_\eta $ and  $\dot{C}_\lambda = (N/2) \dot{c}_\lambda $. 

The expression in Eqn.~\ref{eq:dotcEta} does not depend on  ${\bf v}_{\rm CM}$  because a) in the 2-particle stress  $\sum_{i=1}^2 v_i^xv_i^y$, all cross terms between ${\bf v}_{\rm CM}$ and ${\bf v}_{\rm rel}$ cancel b) the center-of-mass contribution to the stress is conserved during a collision and c) upon averaging over all directions of ${\bf v}_{\rm CM}$, the cross-correlation between $v_{\rm CM}^xv_{\rm CM}^y$ and $\Delta v_{\rm rel}^xv_{\rm rel}^y$ vanishes. 
When we average over all orientations of $X,Y,Z$ with respect to $x,y,z$ we get (see SI):  
\begin{equation}\label{eq:Delta_stress}
v_{\rm rel}^xv_{\rm rel}^y\left(
\Delta  v_{\rm rel}^xv_{\rm rel}^y\right)=
\frac{1}{15}
v^2_{\rm rel}(0)\left(v^2_{\rm rel}(1)P_2(\cos\theta) - v^2_{\rm rel}(0)\right)\;.
\end{equation}
We can then rewrite $\dot{c}_\eta$ as
\begin{equation}\label{eq:dotcEtaA}
\dot{c}_\eta =\frac{\Gamma m^2}{60}
\langle
v^2_{\rm rel}(0)\left(v^2_{\rm rel}(1)P_2(\cos\theta) - v^2_{\rm rel}(0)\right)
\rangle_c
\;.
\end{equation}

Finally, for the heat flux, we obtain:
\begin{eqnarray}\label{eq:dotcLambda}
\dot{c}_\lambda &=&\Gamma^{(1)}\left\langle
\sum_{i=1}^2  v_i^x \left(\frac{1}{2}mv_i^2+ E_{\rm int}^{(i)}-h_i\right)
\Delta \left[ \sum_{i=1}^2  v_i^x \left(\frac{1}{2}mv_i^2+ E_{\rm int}^{(i)}-h_i\right)\right]\right\rangle_c \;.
\end{eqnarray}
Eqn.~\ref{eq:dotcLambda} for heat transport needs further simplification.  Let us focus on the heat flux itself
\begin{equation}
\sum_{i=1}^2  v_i^x \left(\frac{1}{2}mv_i^2+ E_{\rm int}^{(i)}-h_i\right)
=
v_{\rm CM}^x (E_{\rm tot}^{1,2} -2h)+\frac{1}{2}v_{\rm rel}^x \left(E^{(1)}-E^{(2)} \right)\;,
\end{equation}
where $ E_{\rm tot}^{1,2}$ denotes the total energy of the collision pair (1,2), and $E^{(i)}$ denotes the energy of collision partner $i$. 
Note that in the second half of this expression, the average enthalpy has dropped out as it does not change during the collision. 
Moreover, as $ E_{\rm tot}^{1,2}$ is conserved during a collision, we only need to consider the change during the collision of  the second term (involving $E^{(1)}-E^{(2)} $). 
As the directions of ${\bf v}_{\rm CM}$ and ${\bf v}_{\rm rel}$ are uncorrelated, there is no cross correlation between the terms involving $\hat{{\bf v}}_{\rm CM}$ before the collision and the one involving $\hat{{\bf v}}_{\rm rel}$ after the collision. 
Here $\hat{{\bf a}}$ denotes the unit vector in direction of ${\bf a}$. 
However, even though all vectorial correlations between ${\bf v}_{\rm CM}$ and ${\bf v}_{\rm rel}$ vanish, we will still find scalar correlations.

The terms that we need to consider are of the form
\begin{equation}
\left[\frac{1}{2}v_{\rm rel}^x \left(\frac{1}{2}mv_1^2-\frac{1}{2}mv_2^2
+E_{\rm int}^{(1)}-E_{\rm int}^{(2)}\right)\right] \;.
\end{equation} 
As
\begin{equation}
\frac{1}{2}mv_1^2-\frac{1}{2}mv_2^2 = m {\bf v}_{\rm CM}\cdot {\bf v}_{\rm rel}
\end{equation}
we must compute
\begin{gather*}
\frac{1}{2} v^x_{\rm rel}(0)\left[ m {\bf v}_{\rm CM}\cdot {\bf v}_{\rm rel}(0) +E_{\rm int}^{(1)}(0)-E_{\rm int}^{(2)}(0)
\right]\\
 \times
 \frac{1}{2} v^x_{\rm rel}(0)\left[ m {\bf v}_{\rm CM}\cdot {\bf v}_{\rm rel}(0) +E_{\rm int}^{(1)}(0)-E_{\rm int}^{(2)}(0)\right]
 \end{gather*}
 and
 \begin{gather*}
\frac{1}{2} v^x_{\rm rel}(0)\left[ m {\bf v}_{\rm CM}\cdot {\bf v}_{\rm rel}(0) +E_{\rm int}^{(1)}(0)-E_{\rm int}^{(2)}(0)
\right]\\
 \times
 \frac{1}{2} v^x_{\rm rel}(1)\left[ m {\bf v}_{\rm CM}\cdot {\bf v}_{\rm rel}(1) +E_{\rm int}^{(1)}(1)-E_{\rm int}^{(2)}(1)\right]\;,
 \end{gather*}
 where the arguments $(0)$ and $(1)$ denote the states before and after the collision respectively. 
The important point to note is that these expressions still depend of ${\bf v}_{\rm CM}$. 
However, terms linear in ${\bf v}_{\rm CM}$ will vanish upon averaging over orientations. 
We first average 
\begin{gather*}
\frac{1}{2} v^x_{\rm rel}(0)\left[ m {\bf v}_{\rm CM}\cdot {\bf v}_{\rm rel}(0) \right]\\
 \times
 \frac{1}{2} v^x_{\rm rel}(0)\left[ m {\bf v}_{\rm CM}\cdot {\bf v}_{\rm rel}(0)\right]
 \end{gather*}
over a Maxwell distribution of ${\bf v}_{\rm CM}$:
\begin{equation}
\frac{1}{3} \frac{m^2}{4} v_{\rm rel}^2(0)
\left\langle [{\bf v}_{\rm CM}\cdot {\bf v}_{\rm rel}(0) ]^2 \right\rangle_{V_{\rm CM}} 
 =   \frac{m^2}{12} v^2_{\rm rel}(0) v_{\rm rel}^2(0) \frac{k_BT}{ 2m}
  =   \frac{mk_BT}{ 24} v^4_{\rm rel}(0) 
\end{equation}
Similarly
\begin{equation}
\frac{m^2}{ 12} {\bf v}_{\rm rel}(0)\cdot {\bf v}_{\rm rel}(1) 
\left\langle [{\bf v}_{\rm CM}\cdot {\bf v}_{\rm rel}(0) ][{\bf v}_{\rm CM}\cdot {\bf v}_{\rm rel}(1) ] \right\rangle_{V_{\rm CM}}   =   \frac{mk_BT }{ 24} v^2_{\rm rel}(0) v^2_{\rm rel}(1)\cos^2\theta \;.
\end{equation}
The above result follows (for instance) from the spherical-harmonics addition theorem.
Hence, gathering all terms, we get
\begin{eqnarray}\label{eq:jlambda_Delta_jlambda}
j_\lambda(0)\Delta j_\lambda&\equiv& \sum_{i=1}^2  v_i^x \left(\frac{1}{2}mv_i^2+ E_{\rm int}^{(i)}\right)
\Delta \left[ \sum_{i=1}^2  v_i^x \left(\frac{1}{2}mv_i^2+ E_{\rm int}^{(i)}\right)\right]\nonumber\\
&=&\frac{1}{12}\left[ - v^2_{\rm rel}(0)[E_{\rm int}^{(1)}(0)-E_{\rm int}^{(2)}(0)]^2\right.\nonumber\\
&+&v_{\rm rel}(0) v_{\rm rel}(1) \cos\theta [E_{\rm int}^{(1)}(0)-E_{\rm int}^{(2)}(0)]
[E_{\rm int}^{(1)}(1)-E_{\rm int}^{(2)}(1)]\nonumber \\
&+&\left.
\frac{mk_BT}{2} v^2_{\rm rel}(0) \left[v^2_{\rm rel}(1)\cos^2\theta - v^2_{\rm rel}(0)\right]\right]
\end{eqnarray}
which can be written as
\begin{eqnarray}\label{eq:dotcLambda2}
\dot{c}_\lambda &=&\frac{\Gamma}{12}\left[\left\langle - v^2_{\rm rel}(0)[E_{\rm int}^{(1)}(0)-E_{\rm int}^{(2)}(0)]^2\right\rangle_c\right.\nonumber\\
&+&\left\langle v_{\rm rel}(0) v_{\rm rel}(1) \cos\theta [E_{\rm int}^{(1)}(0)-E_{\rm int}^{(2)}(0)]
[E_{\rm int}^{(1)}(1)-E_{\rm int}^{(2)}(1)]\right\rangle_c\nonumber \\
&+&\left.\left\langle
\frac{mk_BT}{2} v^2_{\rm rel}(0) \left[v^2_{\rm rel}(1)\cos^2\theta - v^2_{\rm rel}(0)\right]\right\rangle_c\right]
\end{eqnarray}
Returning now to the Green-Kubo expressions,
\begin{equation}
\int_0^\infty dt\; C_\alpha(t)=-<j_\alpha^2> \left(\frac{<j_\alpha^2>}{ \dot{C}_\alpha(0+)}\right)\;,
\end{equation}
we need to evaluate in the case of diffusion:
\begin{equation}
-<v^2> \left(\frac{<v^2>}{ \dot{c}_D(0+)}\right)\;,
\end{equation}
and
\begin{equation}
\int_0^\infty dt\; C_D(t)=-<v^2> \left(\frac{<v^2>}{ \dot{c}_D(0+)}\right)\;,
\end{equation}
For $\eta$ and $\lambda$ we also first need expressions for $<j_\alpha^2>/\dot{c}_\alpha(0+)$.

Below we give these expressions for $D$, $\eta$ and $\lambda$:
\begin{enumerate}
\item [For D:]
\begin{equation}
 \frac{<v_x^2>^2}{ \dot{C}_D(0+)}= \frac{(k_BT/m)^2}{ \dot{C}_D(0+)}
\end{equation}
where $ \dot{C}_D(0+)$ is given by Eqn.~\ref{eq:dotcD}.
\item[For $\eta$:]
\begin{equation}
\frac{\langle(mv_xv_y)^2\rangle}{ \dot{c}_\eta(0+)}= \frac{(k_BT)^2}{ \dot{c}_\eta(0+)}
\end{equation}
\item[For $\lambda$:]
\begin{eqnarray}\label{eq:jq2}
\lefteqn{\frac{\langle[v_x((1/2)mv^2+(E_{\rm int}-h)]^2\rangle}{ \dot{c}_\lambda(0+)}}&&\nonumber\\
&=& \frac{k_BT}{ m}\frac{\left[<(E_{\rm int}-h)^2>  + 5<(E_{\rm int}-h)> k_BT +(35/4)(k_BT)^2\right]
}{ \dot{c}_\lambda(0+)}\nonumber\\
&=& \frac{k_BT}{m}\frac{\left[k_BT^2C_V^{\rm int} +5/2(k_BT)^2\right]
}{ \dot{c}_\lambda(0+)}
\end{eqnarray}
\end{enumerate}
where we have use the fact that $h$=$E_{\rm int}$+(5/2)$k_BT$ and $<(E_{\rm int}-<E_{\rm int}>)^2>$=
$k_BT^2C_V^{\rm int}$.
To obtain the expressions for the transport properties, we must correct for the fact that, for the stress and the heat flux (but not for the self-diffusion),  $\dot{c}_\alpha$ refers to the change in the current due to  {\em two} particles.
The final expressions are then:
\begin{eqnarray}\label{eq:final_expressions}
\hspace{1cm}D &=&- \frac{(k_BT/m)^2}{ \dot{C}_D(0+)}\\
\eta &=& -2\rho k_BT \frac{(k_BT)^2}{ \dot{c}_\eta(0+)}\\
\lambda &=& -\frac{2\rho}{k_BT^2}\left(\frac{k_BT}{ m}\right)^2\\
&\times&\frac{\left[k_BT^2C_V^{\rm int} +5/2(k_BT)^2\right]^2
}{ \dot{c}_\lambda(0+)}\nonumber
\end{eqnarray}
In what follows, we focus on these three transport properties, although others can be computed as well. 
Examples are given in the SI.
\subsection{Slowly relaxing internal degrees of freedom}
As is clear from Eqns.~\ref{eq:dotcLambda} and S-27  (but implicitly also from the expressions for the other transport properties) energy exchange with internal degrees of freedom affects the transport properties. 
If these degrees of freedom can be described by classical mechanics, there is  no need for additional discussion, unless they are weakly coupled to the collision dynamics. 
However, for polyatomic molecules, many vibrational modes will have frequencies $\nu$ such that the thermal energy $k_BT$ is (much)  less than $h\nu$ (where $h$ is Planck's constant).
There are two reasons why such modes should not be described by classical mechanics: first of all, the heat capacity of quantized modes is less than the corresponding classical value, and secondly, it would be incorrect to describe the collision dynamics of quantum modes with classical mechanics.
Nevertheless, to a good approximation we can still use the framework described above, provided that the ``hard'' (e.g. quantized) modes relax slowly.
We can then make the approximation that the energy stored in these internal degrees of freedom does not relax appreciably during a single collision (although, of course,  all modes will thermalize eventually).   

To analyze the effect of slowly relaxing degrees of freedom, we consider Eqn.~\ref{eq:jlambda_Delta_jlambda}. 
As ``hard'' internal modes are assumed to relax slowly, we can assume that for those modes $[E_{\rm int}^{(1)}-E_{\rm int}^{(2)}]$ is unchanged during a collision.

We consider the general case where some internal modes are ``hard'' and some are ``soft'' in Eqn.~\ref{eq:dotcLambda_hard_soft} at the end of this section.
However,  here we first consider a simpler case, where {\em all} internal modes are assumed to be hard, because this limit leads an interesting approximation for the thermal conductivity.  

If the internal energy is constant during a collision, we have:
\begin{eqnarray}\label{eq:ModifiedEucken}
j_\lambda(0)\Delta j_\lambda
&\approx&\frac{1}{12}\left[ - 
[E_{\rm int}^{(1)}-E_{\rm int}^{(2)}]^2
v_{\rm rel}(0)\left(v_{\rm rel}(1) \cos\theta -v_{\rm rel}(0)\right)
\right.\nonumber\\
&+&\left.
\frac{mk_BT}{2} v^2_{\rm rel}(0) \left[v^2_{\rm rel}(1)\cos^2\theta - v^2_{\rm rel}(0)\right]\right]
\end{eqnarray}
In what follows, we consider the limiting case that the internal energy is completely independent of the translational motion before and after the collision~\footnote{Note that this is a stronger assumption than stating that the internal energy does not change during a collision: even if the internal energy is unchanged, the conformation of a molecule may depend on its internal energy, and this would affect the change in the translational velocity during a collision.}. In that case, we can perform the averaging over internal energies independent of the averaging over translational motion: \begin{eqnarray}\label{eq:ModifiedEucken2}
j_\lambda(0)\Delta j_\lambda
&=&\frac{1}{12}\left[ 
[\langle E_{\rm int}^{2}\rangle-\langle E_{\rm int}\rangle^{2}]
v_{\rm rel}(0)\left(v_{\rm rel}(1) \cos\theta -v_{\rm rel}(0)\right)
\right.\nonumber\\
&+&\left.
\frac{mk_BT}{2} v^2_{\rm rel}(0) \left[v^2_{\rm rel}(1)\cos^2\theta - v^2_{\rm rel}(0)\right]\right]\nonumber\\
&=&\frac{1}{12}\left[ 
[C_V^{\rm int}k_BT^2]
v_{\rm rel}(0)\left(v_{\rm rel}(1) \cos\theta -v_{\rm rel}(0)\right)
\right.\\
&+&\left.
\frac{mk_BT}{3} v^2_{\rm rel}(0) \left[\left(v^2_{\rm rel}(1)P_2(\cos\theta)  - v^2_{\rm rel}(0)\right) 
+(1/2) \left(v^2_{\rm rel}(1)-v^2_{\rm rel}(0)\right)
\right]\right]\;,\nonumber
\end{eqnarray}
where we have used the relation between the variance in the internal energy and the heat capacity due to the internal degrees of freedom.  
Strictly speaking, we could have replaced $v_{\rm rel}(1)$ in Eqn.~\ref{eq:ModifiedEucken2} by $v_{\rm rel}(0)$ (because we now assume elastic collisions). We keep the distinction to make it easier to compare the expression for the thermal conductivity with the ones we obtained earlier for the diffusivity and the viscosity. 
We can now write:
\begin{eqnarray}\label{eq:dotcLambda3}
\dot{c}_\lambda &=&\frac{\Gamma}{12}\left[
[C_V^{\rm int}k_BT^2]
\left\langle v_{\rm rel}(0)\left(v_{\rm rel}(1) \cos\theta -v_{\rm rel}(0)\right)\right\rangle_c
\right.\\
&+&\left.
\frac{mk_BT}{ 3} \left\langle v^2_{\rm rel}(0) \left[\left(v^2_{\rm rel}(1)P_2(\cos\theta)  - v^2_{\rm rel}(0)\right) 
+(1/2) \left(v^2_{\rm rel}(1)-v^2_{\rm rel}(0)\right)
\right]\right\rangle_c\right]
\end{eqnarray}

Using Eqns.~\ref{eq:dotcDA} and \ref{eq:dotcEtaA}, we can write
\begin{equation}\label{eq:dotcDB}
\langle
v_{\rm rel}(0)\left[v_{\rm rel}(1)\cos\theta - v_{\rm rel}(0)\right]
\rangle_c = \frac{12 \dot{C}_D}{\Gamma}\;,
\end{equation}
\begin{equation}\label{eq:dotcEtaB}
\langle
v^2_{\rm rel}(0)\left(v^2_{\rm rel}(1)P_2(\cos\theta) - v^2_{\rm rel}(0)\right)
\rangle_c = \frac{60\dot{c}_\eta} {\Gamma m^2}
\;.
\end{equation}
Then, using
\begin{eqnarray*}\label{eq:final_expressions2}
\hspace{1cm}D &=&- \frac{(k_BT/m)^2}{ \dot{C}_D(0+)}\;,\\
\eta &=& -2\rho k_BT \frac{(k_BT)^2}{ \dot{c}_\eta(0+)}\;,\nonumber
\end{eqnarray*}
we have
\begin{equation}\label{eq:dotcDC}
\langle
v_{\rm rel}(0)\left[v_{\rm rel}(1)\cos\theta - v_{\rm rel}(0)\right]
\rangle_c = -\frac{12 (k_BT/m)^2}{\Gamma D}
\end{equation}
and
\begin{equation}\label{eq:dotcEtaC}
\langle
v^2_{\rm rel}(0)\left(v^2_{\rm rel}(1)P_2(\cos\theta) - v^2_{\rm rel}(0)\right)
\rangle_c = -\frac{120\rho (k_BT)^3}{\Gamma m^2\eta} 
\;.
\end{equation}

we obtain
and hence
\begin{eqnarray}\label{eq:dotcLambda4}
\dot{c}_\lambda &=&\frac{-\Gamma}{ 12}\left[
[C_V^{\rm int}k_BT^2]
\frac{12 (k_BT/m)^2}{\Gamma D}\right.\\
&+&\left. \frac{mk_BT}{ 3} \left(\frac{120\rho (k_BT)^3}{\Gamma m^2\eta} 
+(1/2)\left\langle v^2_{\rm rel}(0)(v^2_{\rm rel}(0)-v^2_{\rm rel}(1))\right\rangle_c
\right)\right]
\end{eqnarray}
The expression for the thermal conductivity then follows from Eqn.~\ref{eq:final_expressions}
\begin{eqnarray}
\lambda &=& \frac{2\rho}{ k_BT^2}\left(\frac{k_BT }{ m}\right)^2\\
&\times&\frac{\left[k_BT^2C_V^{\rm int} +5/2(k_BT)^2\right]^2
}{
\frac{1}{12}\left[
12[C_V^{\rm int}/k_B]
\frac{(k_BT)^4}{ m^2D}
+\frac{mk_BT}{ 3} \left[\frac{120\rho (k_BT)^3}{ m^2\eta} 
+(1/2)\left\langle v^2_{\rm rel}(0)(v^2_{\rm rel}(0)-v^2_{\rm rel}(1))\right\rangle_c/\Gamma
\right]\right]}\nonumber
\end{eqnarray}

We can simplify this expression, using the fact that we have assumed that the internal energy does not change during a collision:

\begin{equation}\label{eq:Mod_Eucken}
\lambda = \frac{\left[C_V^{\rm int}/k_B +5/2\right]^2
}{ 
\left[
\frac{C_V^{\rm int}/k_B}{ 2k_B\rho D}
+\frac{5m}{ 3k_B\eta} 
\right]}=
\frac{R \eta}{M}
\frac{\left[C_V^{\rm int}/k_B +5/2\right]^2
}{ 
\left[
\left(C_V^{\rm int}/k_B\right) {\rm Sc}/2 +(5/3)
\right]}\;,
\end{equation}
where ${\rm Sc}\equiv\eta/(\rho m D)$ is the Schmidt number, $R$ is the gas constant, and $M$ is the molecular weight (in kg/mol) of the molecules. 
Eqn.~\ref{eq:Mod_Eucken}  is similar in spirit, but different in form from the Eucken relation and its many variants~\cite{cha701} . 
As Eqn.~\ref{eq:Mod_Eucken} ignores all relaxation of the internal energy, it will typically overestimate the thermal conductivity.

There are many versions of the Eucken relation: all aim to account for the different relaxation rates of translation, rotation and internal vibration. 
They do so with varying degree of success.
Eqn.~\ref{eq:Mod_Eucken} above is also approximate, and not necessarily particularly accurate.
However, it constitutes an interesting limiting case where internal degrees of freedom do not relax at all on the timescales where velocity and stress decay and, as a consequence, it probably yields an upper limit to the true thermal conductivity. 
A widely used version of the Eucken formula is~\cite{cha701,bar591}:
\begin{eqnarray}\label{eq:Eucken_corr}
\lambda &=& (R\eta/M)  \left[ 15/4 + \frac{C_V^{int}}{{\rm Sc}\;R}\right]\;.
\end{eqnarray}
An advantage of expressions such as Eqns.~\ref{eq:Mod_Eucken} and \ref{eq:Eucken_corr} is that they make it possible to account for strongly quantized internal degrees, as these enter only through the intra-molecular contribution to the heat capacity.
We will use Eqn.~\ref{eq:Eucken_corr} to estimate $\lambda$ of n-octane form the computed values of $\eta$ and $\rho D$, and the tabulated thermal conductivity~\cite{kle101} . 
For atomic gases (no internal degrees of freedom) for which $C_V$=$(3/2) k_B$,  Eqn.~\ref{eq:Mod_Eucken} reduces to
\begin{equation}
\lambda = \frac{(25/4)}
{\frac{5m}{ 2C_V\eta} }
=(5/2) (C_V/m)\eta \;,
\end{equation}
which is a well-known result of the Chapman-Enskog theory.

The assumption that the internal energy does not change during a collision is better justified for internal vibrations that are strongly quantized, than for rotations and low-frequency vibrations.

As mentioned above, the assumption that {\em all} internal modes relax slowly is not essential and can easily be relaxed, yielding
\begin{eqnarray}\label{eq:dotcLambda_hard_soft}
\dot{c}_\lambda &=&\frac{-\Gamma}{ 12}\left[
C_V^{\rm IH}k_BT^2
\frac{12 (k_BT/m)^2}{\Gamma D}\right.\\
&+&
\left\langle - v^2_{\rm rel}(0)[E_{\rm IS}^{(1)}(0)-E_{\rm IS}^{(2)}(0)]^2\right\rangle_c\nonumber\\
&+&\left\langle v_{\rm rel}(0) v_{\rm rel}(1) \cos\theta [E_{\rm IS}^{(1)}(0)-E_{\rm IS}^{(2)}(0)]
[E_{\rm IS}^{(1)}(1)-E_{\rm IS}^{(2)}(1)]\right\rangle_c\nonumber \\
&+&\left.\left\langle
\frac{mk_BT}{2} v^2_{\rm rel}(0) \left[v^2_{\rm rel}(1)\cos^2\theta - v^2_{\rm rel}(0)\right]\right\rangle_c\right]
\end{eqnarray}
where the superscript IH refers to the hard (slowly relaxing) internal modes, and subscript IS refers to the soft, rapidly relaxing modes. Eqn.~\ref{eq:dotcLambda_hard_soft} can be used in simulations, although we have not done so in this paper. 
\subsection{BGK approximation as a lower bound}\label{sec:BGK_bound}
To get an estimate of the effect of memory effects  that are ignored in the BGK approximation, it is useful to start from our expression for the initial average decay rate of the correlation function, and then consider corrections.

Let us therefore consider the correlation function $C_\alpha(t)$, where $J_\alpha$ stands for particle velocity, shear stress or heat flux.

The BGK approximation implies that we approximate the correlation function $C_\alpha(t)$ with a single exponential
\begin{equation}
C_\alpha(t)\approx C_\alpha(0)e^{-t/\tau}\;,
\end{equation}
with
\begin{equation}
1/\tau=
\frac{{\dot C}_\alpha(t=0+)}{C_\alpha(0)};.
\end{equation}

To analyze the more general case, we make use of the fact that the decay of correlations in a gas at infinite dilution is a Markov process, meaning that the rate of change of a flux $j_\alpha(t)$ depends only on the (complete) set of dynamical variables $Y$ that characterize $j_\alpha$. We can then  write
\begin{equation}\label{eq:Liouville}
\dot{j}_\alpha(t) \equiv -L j_\alpha \;, 
\end{equation}
which defines the operator $L$. 
The  formal solution of Eqn.~\ref{eq:Liouville} is
\begin{equation}\label{eq:Liouville2}
j_\alpha(t) = e^{ -L t} j_\alpha(0)  \;, 
\end{equation}
and hence
\begin{equation}\label{eq:Liouville3}
C_\alpha(t)  = \langle j_\alpha(0)e^{ -L t} j_\alpha(0)\rangle =\langle j_\alpha^2\rangle
\frac{\langle j_\alpha e^{ -L t}j_\alpha\rangle}{\langle j_\alpha^2\rangle} = \langle j_\alpha^2\rangle\langle\langle e^{ -L t}\rangle\rangle
\;, 
\end{equation}
where 
\[
\langle\langle\cdots\rangle\rangle \equiv \frac{\langle j_\alpha \cdots j_\alpha\rangle}{\langle j_\alpha^2\rangle} \;.
\]
Of course, the solution of Eqn.~\ref{eq:Liouville3} would require solving the full kinetic equation, which is the core problem of kinetic theory.
But, even without solving Eqn.~\ref{eq:Liouville3}, we can  make a general statement about the BGK approximation.
Note that, in the language of Eqn~\ref{eq:Liouville2}, the BGK approximation can be written as
\begin{equation}\label{eq:Liouville3b}
C^{\rm BGK}_\alpha(t) =  \langle j_\alpha^2\rangle
e^{ -\langle\langle L\rangle\rangle t}   \;,
\end{equation}
where, as before, 
\[
\langle\langle L\rangle\rangle\equiv 
\frac{\langle j_\alpha L j_\alpha\rangle}{\langle j_\alpha^2\rangle}\;.
\]
Note that, as $t\rightarrow 0+$, the full correlation function approaches the BGK expression (as it should):
\[
\lim_{t\rightarrow 0+}\langle\langle  e^{ -L t}\rangle\rangle = 1- \langle\langle L\rangle\rangle t= \lim_{t\rightarrow 0+}e^{ -\langle\langle L\rangle\rangle t} \;.
\]
Now we can use the fact that the exponential is a convex function. 
{\color{red}Provided the exponential is averaged over a non-negative weight function,} Jensen's inequality~\cite{jen061} applies to  Eqns.~\ref{eq:Liouville3} and \ref{eq:Liouville3b}, implying that 
\begin{equation}\label{eq:inequality}
\langle\langle  e^{ -L t}\rangle\rangle \ge e^{-\langle\langle L\rangle\rangle t} \;. 
\end{equation}
This inequality implies that the BGK estimate of a transport property must necessarily be a lower bound to the true transport property, provided that the averaging is over a distribution with non-negative weights.
\section{Simulations}
We have tested the expression (eqn.~\ref{eq:final_expressions}) for $D$, $\eta$ and $\lambda$ for a number of simple cases where the Chapman-Enskog results are known. In particular, we consider hard spheres and a (truncated) Lennard-Jones model as prototypical examples of particles with no internal degrees of freedom.

After that, we consider one of the few examples of a model of particles with internal degrees of freedom. 
We chose the rough hard-sphere model, which has been studied by several authors (see refs.~\citenum{pid221}, \citenum{con651}, \citenum{mcc661} and \citenum{cha701}). 

\subsection{Hard Spheres}\label{sec:HS}
The transport coefficients of a dilute gas of smooth hard spheres were computed using the current approach and compared with the analytical results obtained with the Chapman-Enskog approach,  with and without  higher-order corrections.

For the simulations, we wrote a (trivial) 2-particle, event-driven hard-sphere MD code. 
The samples from a Maxwell distribution of relative velocities were drawn using the Box-Muller method.
To integrate over the impact parameters between 0 and $r_c$, we used a 10-point Gauss-Legendre (GL) quadrature. 
For every impact parameter, we ran 10$^5$ trajectories.
We note that the BGK results are in good agreement with the lowest-order Chapman-Enskog results. 
We find the same for the case of rough hard spheres (discussed below).
This finding suggests that the BGK approximation and the lowest-order Chapman-Enskog expression may be equivalent, at least in some cases. 
We did not explore this.

The comparison of the numerical (BGK) results are compared with the analytical results in Table~\ref{tab:hard_spheres}
\begin{table}
\begin{tabular}{|l|c|c|c|}
\hline
Transport Coefficient &C-E  & C-E - higher order & Simulations\\
\hline
$\rho D$ & 0.2116 & 0.2149&0.2119(3)\\
$\eta$ & 0.1763 & 0.1791&0.175(1) \\
$\lambda$ & 0.6622 &0.6717&0.661(2)\\
\hline
\end{tabular}
\caption{\label{tab:hard_spheres} Transport coefficients of hard-sphere gas. 
The column labeled C-E gives the lowest order, Chapman-Enskog analytical results~\cite{cha701} . 
The column labeled ``C-E - higher-order'' gives the best available analytical expression~\cite{con651} . 
The last column presents the simulation results. The error estimates (1$\sigma$) are indicated between brackets.}
\end{table}


\subsection{Lennard-Jones gas}\label{sec:LJ}
To compute the transport coefficients of a dilute LJ gas, we had to perform a large number of binary collisions. In order to make the approach tractable for other users, we used the LAMMPS program package to carry out these simulations. To be precise, we prepared the system as if it were a normal MD run, except that individual  collision pairs (initially placed at a distance slightly larger than $r_c$), were far removed from all other collision pairs.
We then created the neighborlist only once and never updated it.
For every collision pair $ij$, the neighbor list contains only one entry. 
Hence, the simulations are cheap and many collisions (100 000) could be run in parallel. 
The advantage of this procedure is that it makes the overhead for starting and ending the runs negligible compared with the computational cost of the simulation itself.

The samples from a Maxwell distribution of relative velocities were drawn using the Box-Muller method.

Collision runs were stopped when particles were no longer interacting and moving away from each other. 
We used a time step  $\Delta t$ = 0.001 (in reduced units). 
The Lennard-Jones potential was truncated  at a cutoff radius $r_c$ = 2.5 $\sigma$. The force was linearly shifted, such that it vanishes continuously at $r_c$. 
As before, we use a 10-point GL quadrature to integrate over impact parameters, and we use 10$^5$ trajectories per impact parameter.

\begin{figure*}[htb]
\centering
\includegraphics[width=\linewidth]{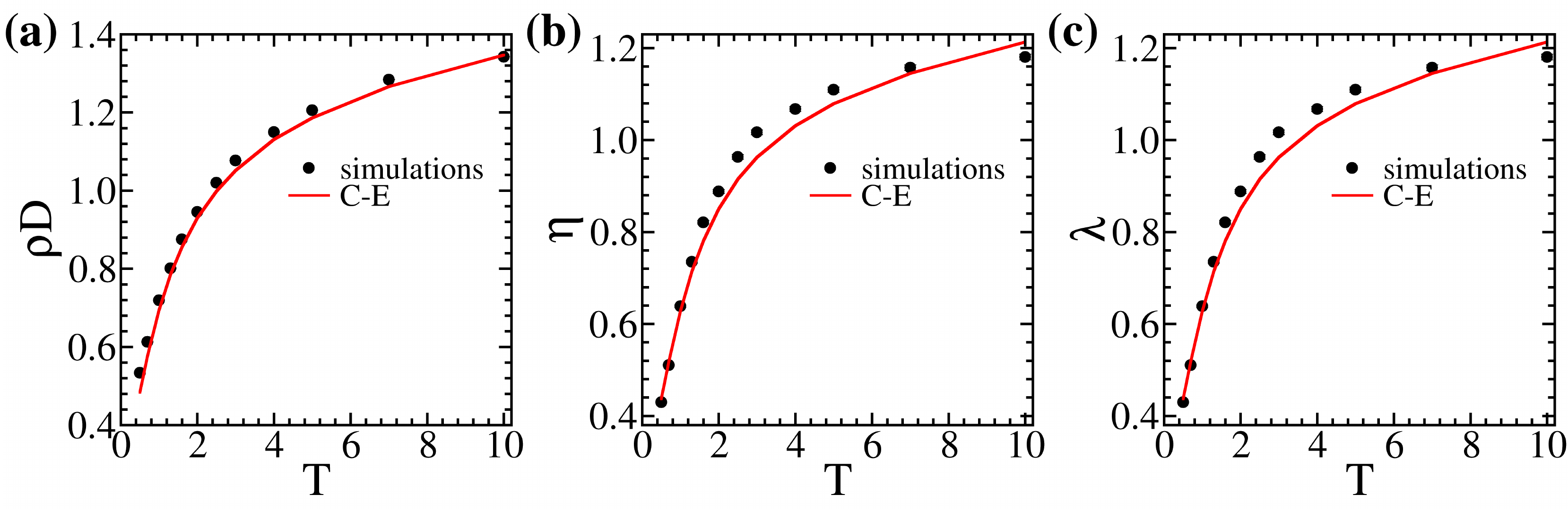}
\caption{Temperature dependence of (a) the diffusivity D, multiplied with the number density $\rho$, (b) the viscosity $\eta$ and (c) the thermal conductivity $\lambda$ of a dilute Lennard-Jones gas.   
The drawn curves represent the Chapman-Enskog predictions (see re.~\citenum{cha701}), the filled circles represent the simulation results obtained using the BGK approximation. 
}
\label{fig:LJ}
\end{figure*}
\clearpage
\subsection{Rough Hard Spheres}\label{sec:RHS}
The rough hard sphere model is one of the simplest models for a molecule that can exchange both translational and rotational kinetic energy in a collision (the other simple model is the loaded hard sphere model). 
The model was presumably introduced by G.H. Bryan in 1894 (although the references in ref.~\citenum{pid221} and \citenum{kra121} are different and hard to trace). 
As in the case of hard spheres, we used an event-driven, binary collision code to compute $\rho D$, $\eta$ and $\lambda$ using Eqn.~\ref{eq:final_expressions}.
The samples from a Maxwell distribution of relative velocities and initial angular momenta were drawn using the Box-Muller method and we
used a 10-point GL quadrature to integrate over impact parameters, and we use 10$^5$ trajectories per impact parameter. 
For the rough hard-sphere model (RHS), the lowest order Chapman-Enskog approach~\cite{pid221} and higher order expressions~\cite{con651, kra121} are known. 
Hence, we use this model as a test of our approach for a molecule with internal energy. 
\begin{figure*}[htb]
\centering
\includegraphics[width=\linewidth]{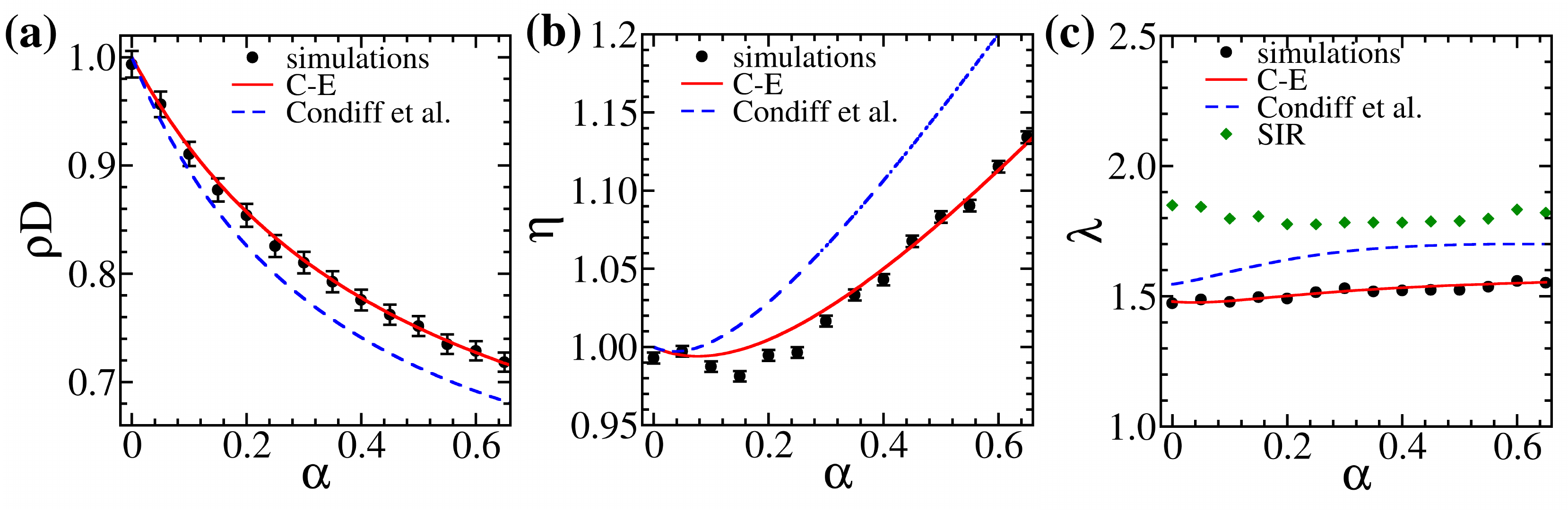}
\caption{Dependence of the transport coefficients of Rough Hard Spheres, as a function of the dimensionless moment of inertia $\alpha$, defined as $\alpha\equiv 4I/(m\sigma^2)$, where $I$ is the moment of inertia of the sphere, $m$ its mass, and $\sigma$ its diameter. All transport coefficients have been divided by the Chapman-Enskog value for hard spheres. Note that the thermal conductivity of rough hard spheres with vanishing moment of inertia is not the same as that of hard spheres, because the latter have no rotational kinetic energy. The figure shows: (a) the diffusivity $D$, multiplied with the number density $\rho$, (b) the viscosity $\eta$ and (c) the thermal conductivity $\lambda$. 
The drawn curves represent the Chapman-Enskog predictions (see~\citenum{cha701}) and the dashed curves correspond to the higher-order analytical results of ref.~\citenum{con651} - or to be more precise, the corrected expressions given in ref.~\citenum{kra121}. The filled circles represent the simulation results obtained  using the BGK approximation. The filled diamonds in (c) correspond to the Slow Internal Relaxation (SIR) approximation (Eqn.~\ref{eq:Mod_Eucken}).}
\label{fig:RHS}
\end{figure*}
The results of this comparison are shown in Fig.~\ref{fig:RHS}. 
As is clear from the figure, our numerical results agree well with the lowest-order Chapman-Enskog results, but may differ by up to 10 \% from the more refined analytical expressions. 
This discrepancy is probably due to the errors introduced by the BGK approximation, although, as we argue below, there may also be problems with the higher-order
truncation of the Chapman-Enskog expansion.  

Fig.~\ref{fig:RHS}(c) also shows the Slow Internal Relaxation (SIR) approximation for $\lambda$.
{\color{red}The SIR approximation yields a higher estimate for $\lambda$ than BGK, consistent with Eqn.~\ref{eq:inequality} discussed below.  
The  result of ref.~\citenum{con651} for the diffusivity is  problematic because Eqn.~\ref{eq:inequality} suggests that the BGK equation should yield a lower bound to $D$, $\eta$ and $\lambda$.}
As is clear from Fig.~\ref{fig:RHS}a, the  result of ref.~\citenum{con651} for $\rho D$ does not satisfy this inequality.
\subsection{Nitrogen gas}\label{sec:N2}
Using the approach described above, we compute $\rho D$, $\eta$ and $\lambda$ for a model of nitrogen at temperatures between 233.15 K and 573.15 K. 
For the sake of comparison with the MD simulations (see below), we used the same 2-center Lennard-Jones model of nitrogen  with a rigid bond constraint as used in ref.~\citenum{lee141} (be it that we assumed  a cutoff distance of  0.8 nm, rather than the somewhat excessive 8 nm mentioned in ref.~\cite{lee141}). 
For the trajectory calculations, we used a time step of 10$^{-15}$ s. 
We used 10 000 collision pairs per run and, to carry out the binary collision calculations efficiently in LAMMPS,  we did not update the neighbor list.

\begin{figure*}[htb]
\centering
\includegraphics[width=\linewidth]{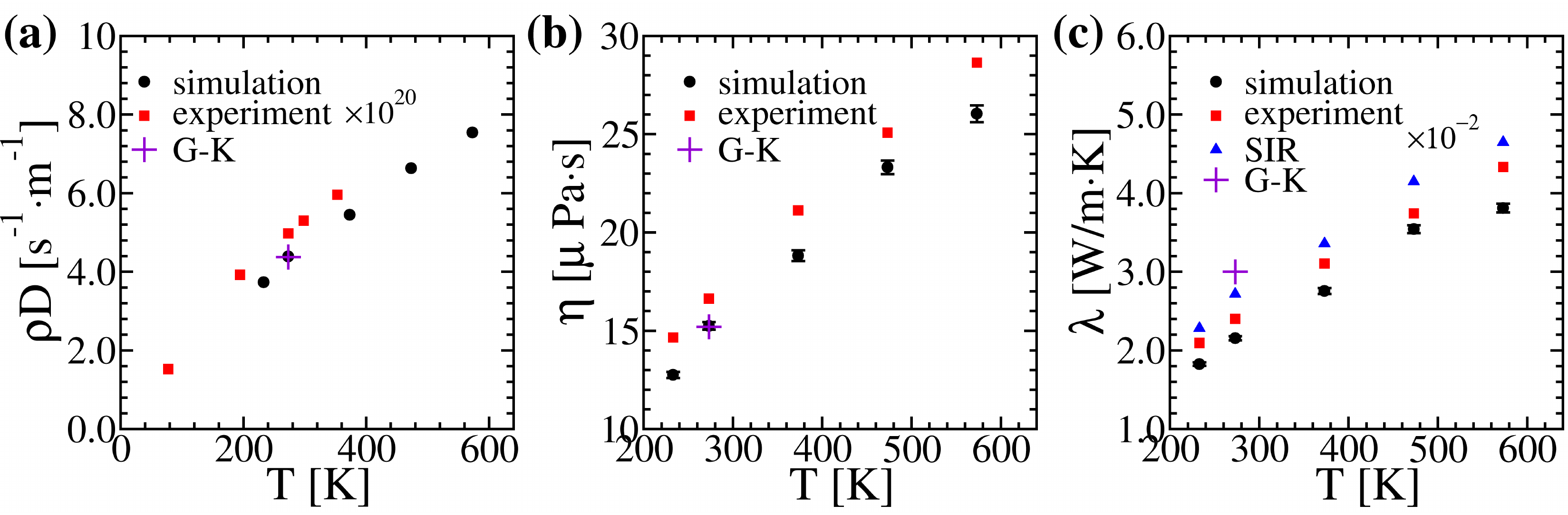}
\caption{Temperature dependence of (a) the diffusivity D, multiplied with the number density $\rho$, (b) the viscosity $\eta$ and (c) the thermal conductivity $\lambda$ of nitrogen gas ($N_2$) at infinite dilution.
We compare our numerical results with experimental data for nitrogen  gas~\cite{kle101} (filled squares). 
The triangles in panel (c) correspond to the Slow Internal Relaxation (SIR) approximation (Eqn.~\ref{eq:Mod_Eucken}).
The crosses indicate the result of brute-force MD simulations for nitrogen at atmospheric pressure and 273 K.} 
\label{fig:N2}
\end{figure*}

The samples from a Maxwell distribution of relative velocities were drawn using the Box-Muller method. 
We used a short, low-density MD run to generate a sample of randomly oriented, non-interacting molecules  with an equilibrated distribution of angular velocities. 
We used a 10-point GL quadrature to integrate over impact parameters, and we use 10$^4$ trajectories per impact parameter. 
The results are shown in figure~\ref{fig:N2}. In the same figure, we also show the corresponding experimental results. The values for $\eta$ and $\lambda$ were taken from the reference data of ref.~\citenum{kle101}. The self-diffusion constant of N$_2$ was reported in ref.~\citenum{win501}(to be precise, the data apply to $^{15}$N-$^{14}$N). 
The discrepancy between the BGK simulations and the experiments is not surprising, as we have used a very simple model of nitrogen.

In addition, we performed low-density MD simulations  performed using LAMMPS~\cite{pli951} . 
As in ref.~\cite{lee141}, the number of particles was $N$= 1728, and  the time step $\Delta$t = 10$^{-14}$s (to facilitate comparison with ref.~\cite{lee141}). To compute the transport properties (in particular, $\eta$ and $\lambda$), rather long simulations were needed. 
We performed 10 simulations of length 10 ns, excluding 4 ns equilibration time. 
The transport properties were computed using the Green-Kubo expressions, but in the case of the diffusivity, we also computed $D$ from the mean-square displacement.

In panel (c) we have also indicated the estimate for $\lambda$ that follows from the approximation given in Eqn.~\ref{eq:Mod_Eucken}, which assumes Slow Internal Relaxation (SIR). 
Note that, whereas the GK estimate for $D$ and $\eta$ agree very well with the BGK approximation, the MD results for $\lambda$ is closer to the SIR approximation. 
Note that the SIR approximation uses the values for $\rho D$ and $\eta$ listed in the table. 

\clearpage
\subsection{Polyatomic molecules}
To illustrate the fact that the present approach can  compute quantities that are completely beyond reach for the Chapman-Enskog theory, we give just one example: the transport properties of dilute $n$-octane vapor at T=300K. 
Again, we used LAMMPS to run many independent binary collisions in parallel. 

For the sake of simplicity, we chose a Lennard-Jones-based, united atom force field (NERD)~\cite{nat981} . 
We do not expect this force field to be particularly accurate.
For the trajectory calculations, we used a time step of 10$^{-15}$ s. 
We carried out the binary collision calculations on 10 000 pairs in parallel, using LAMMPS without a neighbor list update.

The samples from a Maxwell distribution of relative velocities were drawn using the Box-Muller method. 
We used a short, low-density NVT-MD run to generate a sample of well equilibrated, non-interacting molecules.
We used a 10-point GL quadrature to integrate over impact parameters, and we use 10$^4$ trajectories per impact parameter. 
We also computed $\rho D$  and $\eta$ using brute-force MD of 1000 octane molecules (using LAMMPS). 
In both cases, the BGK results are not statistically different from the MD simulation results.
We note, however, that in spite of the fact that we used long MD simulations to compute $\eta$ (40 ns), the statistical error is large ($\sim$ 6\%), as is to be expected.
We did not compute $\lambda$ by MD, because classical simulations do not adequately account for the heat capacity of quantized degrees of freedom.  
Our estimate for $\lambda$, labeled with (*) is based on our numerical results for $\rho D$ and $\eta$, combined with the corrected Eucken expression~\ref{eq:Eucken_corr}, using the known value for the internal heat capacity~\cite{kle101} .
  
All results have been collected in Table~\ref{tab:n-octane}.
The experimental data were taken from ref.~\citenum{kle101}.
As can be seen, our estimate for $\eta$ is in fair  agreement with experiment,  the agreement for $\lambda$ is better, but this may be fortuitous. 
We show this example as an illustration of the fact that, with the present approach, we can produce reasonable estimates for the transport coefficients of poly-atomic gases, based on our knowledge of the molecular force-field. 

\begin{table}\label{tab:octane}
\begin{tabular}{|cl|c|c|c|}
\hline
Transport Coefficient && present results&  MD simulations & Experiment\\
\hline
$\rho$ D &[s$^{-1}$m$^{-1}$] & 3.87(3) 10$^{19}$& 3.30(3) 10$^{19}$& --\\
$\eta$ &[$\mu$Pa s] & 5.73(7)& 5.2(3) & 5.2\\
$\lambda$ &[W m$^{-1}$ s$^{-1}$]& 0.0124(*) & -- &0.012\\
\hline
\end{tabular}
\caption{\label{tab:n-octane}  Transport coefficients of n-octane, computed using Eqn.~\ref{eq:final_expressions}, are compared with experiment (where available). 
Also shown are  the estimates for $\rho D$ and $\eta$ from brute-force MD. 
Our estimate for $\lambda$, labeled with (*) is based on our numerical results for $\rho D$ and $\eta$, combined with the corrected Eucken expression~\ref{eq:Eucken_corr}, using the known value for the internal heat capacity~\cite{kle101} (see text).}
\end{table}

\section{Conclusions}
In this paper we proposed a scheme to compute the transport coefficients of dilute molecular gases. 
The scheme is based on computing the short-time decay of the correlation functions that appear in the relevant Green-Kubo expressions for $D$, $\eta$ and $\lambda$ (the diffusivity, viscosity and thermal conductivity) of the gas. 
The method that we use is approximate, as it assumes that the correlation functions can be approximated by single exponentials. 
This approximation, that is similar in spirit to the BGK approximation in kinetic theory, yields results that are in good agreement with the lowest-order Chapman-Enskog expressions (where these are available).
However, our results deviate by up to 10\% from the results that are obtained by including higher-order terms in the solution of the Boltzmann equation. 

For realistic molecular models with internal degrees of freedom, the analytical techniques used to solve the Boltzmann equation become useless. 
Hence, in these cases, the only alternative is to perform large scale MD simulations on very dilute gases. 
Such simulations are orders-of-magnitude more time consuming than our simulations, and even then suffer from serious statistical errors, as the relevant correlation functions decay very slowly at low densities.
We show that with our approach, we can arrive at reasonable predictions for the transport coefficients of a typical poly-atomic vapor (n-octane).
The approach that we propose can easily be extended to mixtures. As we assume that intermolecular interactions have a finite range, the expressions  for the transport properties given here are  less suited for molecules with long-ranged interactions. However, this limitation can be overcome without too much trouble.

In principle, the technique that we propose could be made more accurate (e.g. by not making the BGK approximation). 
Whilst it would be attractive to explore more rigorous approaches to compute gaseous transport properties, we argue that this should be done in a judicious way because otherwise the method could quickly become quite complex and would run the risk of suffering the same fate as the higher-order schemes to solve the Boltzmann equation: these methods are barely used by the non-expert, as  {\em the perfect is the enemy of the good}.

\begin{suppinfo}
The Supplementary Information describes:
\begin{enumerate}
    \item How the present approach can be extended to the case of long-ranged intermolecular forces between neutral molecules.
    \item How to compute the effect of a collision on the stress tensor.
    \item A summary of relevant published analytical expressions for the transport coefficients.
    \item a brief description of the expression of other transport coefficients
\end{enumerate}
\end{suppinfo}

\section*{Acknowledgements}
This work was started during a stay of DF as Onsager Professor at the NTNU in Trondheim. DF acknowledges the hospitality of Porelab of the NTNU, and is grateful to the Research Council of Norway for its Center of Excellence funding scheme, PoreLab, project no 262644. In addition, DF gratefully acknowledges discussions with Astrid de Wijn, Kim Kristiansen, Dick Bedeaux and Signe Kjelstrup at Porelab, Alex Routh and Giovanni Ciccotti (who has his reservations). 

Last but not least: DF expresses his gratitude to Emily Carter - a long-term friend and an inspiring colleague, if ever there was one. 
\mciteErrorOnUnknownfalse


\bibliography{main}
\pagebreak

\begin{center}
\textbf{\large Supplementary Information\\Using Molecular Simulation to Compute Transport Coefficients of Molecular Gases}
\end{center}
\setcounter{equation}{0}
\setcounter{figure}{0}
\setcounter{table}{0}
\setcounter{page}{1}
\makeatletter
\renewcommand{\theequation}{S\arabic{equation}}
\renewcommand{\thefigure}{S\arabic{figure}}

\section{Long-ranged forces}\label{sec:app-LR}

In the main text of the article, we have assumed that the collision integrals can be truncated for impact parameters larger than $r_c$, the cutoff distance of the potential. 
Whilst this is often adequate, there may be cases where we wish to account for the effect of long-ranged interactions. 
Below, we argue that this is, in fact possible, provided that the intermolecular potential decays faster than $1/r$.

First, we note that collision integrals are of the form
\begin{eqnarray*}
 \dot{c}_\alpha&=& \rho \left\langle j_\alpha(0)\Delta j_{\alpha;c}\right\rangle\\ 
 &=&\rho \int dX_{\rm int}P(X_{\rm int})
 \int_0^{\infty}\pi db^2\int_0^\infty dv_{\rm rel} \;  v_{\rm rel}  P(v_{\rm rel})\\
&\equiv&
\int_0^{\infty}\pi db^2 \avdbl{v_{\rm rel} j_\alpha(0)\Delta j_{\alpha;c}(b, v_{\rm rel})}\;,
\end{eqnarray*}
where the double brackets denote averaging over internal degrees of freedom and over relative velocities.
We can rewrite this integral as:
\begin{equation}\label{eq:dotc}
 \dot{c}_\alpha = \int_0^{\infty}\pi db^2 g(b^2) \frac{\avdbl{v_{\rm rel} j_\alpha(0)\Delta j_{\alpha;c}(b, v_{\rm rel})}}{g(b^2)} \;,
\end{equation}
where $g(b^2)$ is a function that is larger than zero, but otherwise arbitrary.  
We now choose $g(b^2)$ such that it decays to zero as $b^2\rightarrow\infty$, but not faster than the term between double brackets.
The large-$b$ behavior of the term between double brackets depends on the behavior of the intermolecular potential at large distances. 
We now assume the case of practical interest that the leading $r$-dependence of the pair potential at large $r$ goes as $v(r)\sim r^{-m}$. 
Then the intermolecular force decays as $f(r)=\alpha/r^{m+1}$, where $\alpha$ is a measure for the strength of the interaction.
For large enough impact parameters, the deflection angle $\theta$ of the molecular trajectory becomes small and the quantity $j_\alpha(0)\Delta j_{\alpha;c}(b, v_{\rm rel}))$ scales as $\sin^2\theta$. 
In the same limit, we can write
\[
\sin\theta \approx \frac{\Delta p}{p_0}
\]
where $p_0$ denotes the original relative momentum, and $\Delta p$ is its change due to the long-ranged intermolecular forces. 
To lowest order in $\theta$, we can ignore the fact that the magnitude of the relative momentum changes.
Therefore, we can estimate $\Delta p$ by integrating the force along a straight trajectory:
\begin{eqnarray}\label{eq:deltap}
\Delta p &\approx&\int_{-\infty}^{+\infty} dt\; f(r(t))\times \frac{b}{r(t)}\nonumber\\
&=&\alpha\int_{-\infty}^{+\infty} dt\; b\left( \frac{1}{b^2+v_0^2t^2}\right)^{(m+2)/2}\nonumber\\
&=&\frac{\alpha}{v_0b^m}\int_{-\infty}^{+\infty} dx\;  \left(\frac{1}{1+x^2}\right)^{(m+2)/2}\;,
\end{eqnarray}
where $v_0=p_0/\mu$ and $x\equiv v_0t/b$. 
The integral in the last line certainly converges for $m>1$, which, as we show below, is the range of interest.
We are not interested in the numerical value of the integral, just in the fact that it converges.
We then have that
\begin{equation}\label{eq:sintheta1}
\sin\theta \approx \frac{\Delta p}{p_0}\sim  \frac{1}{p_0^2b^m} \;.
\end{equation}
Note that $\sin\theta$ cannot diverge for small $p_0$, as $|\sin\theta|\le 1$.
From Eqn.~\ref{eq:sintheta1} it follows that $j_\alpha(0)\Delta j_{\alpha;c}(b, v_{\rm rel}))$ scales as $b^{-2m}$.
We can now choose a suitable functional form for $g(b^2)$.
Many choices are possible. Here we consider, as an example, the following simple form:
\begin{eqnarray}\label{eq:gb2} 
   g(b^2) &=& 1 \;\;\; \;\;\;\;\;\;  \;\;\;\mbox{for} \;b^2\le r^2_c \nonumber\\
   &=& (r_c/b)^{2\ell} \;\;\;  \mbox{for} \; b^2 > r^2_c 
\end{eqnarray}
We choose $\ell>1$, such that the integral $ \int_0^{\infty}\pi db^2 g(b^2) $ converges. We do this to ensure that we can write the initial slow of the flux-correlation functions as a product of an effective collision frequency and a change per collisions,  as in Eqns. 28, 32 and 35 in the main text.

We note that $\ell\le m$, because $g(b^2)$ must not decay faster to zero than $j_\alpha(0)\Delta j_{\alpha;c}(b, v_{\rm rel}))$. 
This condition implies that $m>1$.  This is not a serious constraint if we only consider  uncharged molecules.
It is useful to split the integral in Eqn.~\ref{eq:dotc} in  the part for $b< r_c$ and $b\ge r_c$:
\begin{eqnarray}\label{eq:integral}
\int_0^{\infty}\pi db^2 g(b^2)  \left[\cdots\right] &=& \int_0^{r_c^2}\pi db^2  \left[\cdots\right] +
\int_{r_c^2}^{\infty}\pi db^2 \left(\frac{r_c^2}{b^2}\right)^\ell  \left[\cdots\right]\nonumber\\
&=& \int_0^{r_c^2}\pi db^2  \left[\cdots\right] +
\frac{\pi}{\ell - 1} \int_0^1d\left(\frac{r_c^2}{b^2}\right)^{\ell-1} \left[\cdots\right]\;,
\end{eqnarray}
where 
\[
 \left[\cdots\right] = \frac{\avdbl{v_{\rm rel} j_\alpha(0)\Delta j_{\alpha;c}(b, v_{\rm rel})}}{g(b^2)} \;,
\]
which is bounded for $r_c\le b<\infty$. 
Both parts of the integral in Eqn.~\ref{eq:integral} can be computed numerically by sampling and quadrature. 
Hence the framework described in the main text can also be applied to long-ranged potentials.
It seems plausible that an optimal choice of $g(b^2)$ would be one where the $b^2$-dependence of $g(b^2)$ would be the same as for $j_\alpha(0)\Delta j_{\alpha;c}(b, v_{\rm rel}))$. However, we did not investigate this. 
\section{Stress correlation}

The full stress tensor is a traceless dyadic. ${\bf \Sigma}$=$m({\bf v}{\bf v}-(1/3){\bf I})$,
\begin{gather}
{\bf \Sigma}=m
\begin{bmatrix}
v_xv_x-1/3& v_xv_y & v_xv_z\\
v_yv_x& v_yv_y-1/3& v_yv_z\\
v_zv_x&v_zv_y& v_zv_z-1/3\\
\end{bmatrix}
\end{gather}
First, we need to compute the orientational average $(v_xv_y)^2$ (or equivalent). Using polar coordinates, we can write
$v_x=v\cos\theta$ and 
$v_y=v \sin\theta\cos\phi$. Hence
\begin{equation}
<(v_xv_y)^2> = \frac{1}{15} v^4\;.
\end{equation}
Rather then working with components, it is better to compute the trace of $\bf\Sigma : \bf \Sigma$, as this quantity is rotationally invariant.
Upon orientational averaging 
\begin{equation}
<(v^2_x-1/3)^2> = \frac{4}{ 45} v^4 \;.
\end{equation}
The diagonal elements of the orientational average of $\bf\Sigma : \bf \Sigma$ are all equal to
\begin{equation}
\Sigma^2_{xx} = <(v^2_x-1/3)^2>+2<(v_xv_y)^2> = \frac{10}{45} v^4
\end{equation}
or
\begin{equation}
<(v_xv_y)^2> = \frac{1}{10} \mbox{Tr} \bf\Sigma : \bf \Sigma \;,
\end{equation}
with $\mbox{Tr} \bf\Sigma : \bf \Sigma$ = $\frac{2}{3} v^4$. 

The advantage of considering the whole matrix $\bf\Sigma$ is that it transforms as an irreducible tensor of rank 2. Hence, if the post-collisional stress is 
\begin{gather}
{\bf \Sigma}'=m
\begin{bmatrix}
{v'}_x{v'}_x-1/3& {v'}_x{v'}_y & {v'}_x{v'}_z\\
{v'}_y{v'}_x& {v'}_y{v'}_y-1/3& {v'}_y{v'}_z\\
{v'}_z{v'}_x&{v'}_z{v'}_y& {v'}_z{v'}_z-1/3\\
\end{bmatrix}
\end{gather}
then
\begin{equation}
\mbox{Tr}{\bf\Sigma : \bf \Sigma'} = \frac{2}{3}v^2{v'}^2
P_2(cos\theta)
\end{equation}
where $\theta$ is the angle between the pre and post collisional velocities.

Using the relation between the trace and the elements of $\bf\Sigma : \bf \Sigma'$, we obtain
\begin{equation}
<{v}_x{v}_y {v'}_x{v'}_y>= \frac{1}{ 15} v^2{v'}^2 P_2(\cos\theta)
\end{equation}
Hence
\begin{equation}
<{v}_x{v}_y {v'}_x{v'}_y>-<({v}_x{v}_y)^2> 
= \frac{1}{15} \left(v^2{v'}^2 P_2(\cos\theta)- v^4\right)\;,
\end{equation}
which is Eqn.~33 in the main text. 

\section{Summary of published analytical expressions for transport coefficients}

It is convenient to make the diffusion coefficient $D$, viscosity $\eta$ and thermal conductivity $\lambda$ of rough hard sphere and Lennard-Jones particles dimensionless by dividing them by the corresponding   first-order Chapman-Enskog approximations for smooth, hard spheres~\cite{cha701} . The smooth hard sphere expressions are: 
\begin{equation}
\rho D=\frac{3}{8\sigma^2}\sqrt{\frac{k_BT}{\pi m}}
\end{equation}
\begin{equation}
\eta=\frac{5}{16\sigma^2}\sqrt{\frac{mk_BT}{\pi}}
\end{equation}
\begin{equation}
\lambda=\frac{75}{64\sigma^2}\sqrt{\frac{k_B^3T}{\pi m}}
\end{equation}
Pidduck~\cite{pid221}, evaluated the first-order Chapman-Enskog approximation  for rough hard spheres:
\begin{equation}
\rho D=\frac{3}{8\sigma^2}\sqrt{\frac{k_BT}{\pi m}}\frac{1+\alpha}{1+2\alpha}
\end{equation}
\begin{equation}
\eta=\frac{5}{16\sigma^2}\sqrt{\frac{mk_BT}{\pi}}\frac{6(1+\alpha^2)}{6+13\alpha}
\end{equation}
\begin{equation}
\lambda=\frac{75}{64\sigma^2}\sqrt{\frac{k_B^3T}{\pi m}}\frac{12(1+\alpha)^2(37+151\alpha+50\alpha^2)}{25(12+75\alpha+101\alpha^2+102\alpha^3)}
\end{equation}
where $\alpha$ is the reduced moment inertia, $\alpha=4I/m\sigma^2$ and $I$ is moment of  inertia, of the rough hard sphere. The value of $\alpha$ varies from zero (all mass at the center) to 2/3 (all mass on the surface). A sphere with uniform mass density correspond has $\alpha=2/5$.

To our knowledge, the highest-order analytical approximation for the transport coefficients of rough hard spheres was given by Condiff, Lu, and Dahler~\cite{con651} (we used the corrected expression of ref.~\citenum{kra121}), using a Sonine polynomial expansion.
The expression obtained by Condiff {\em et al.} may differ by up to 10\% from the lowest order Chapman-Enskog approximation:
\begin{equation}
\rho D=\frac{3}{8\sigma^2}\sqrt{\frac{k_BT}{\pi m}}\frac{1+\alpha}{1+2\alpha}\times\left [1+\frac{\pi\alpha(1+\alpha)}{2(1+2\alpha)(5+9\alpha+8\alpha^2)}\right]^{-1}
\end{equation}
\begin{equation}
\eta=\frac{5}{16\sigma^2}\sqrt{\frac{mk_BT}{\pi}}\frac{2(1+\alpha)^2(3+10\alpha)}{6+33\alpha+35\alpha^2}
\end{equation}
\begin{equation}
\lambda=\frac{75}{64\sigma^2}\sqrt{\frac{k_B^3T}{\pi m}}\frac{4(1+\alpha)(1121+7336\alpha+13449\alpha^2+9490\alpha^3+2000\alpha^4)}{25(116+853\alpha+1707\alpha^2+2266\alpha^3+1360\alpha^4)}
\end{equation}

The first-order Chapman-Enskog approximation for the transport properties of a Lennard-Jones gas is given, for instance, in the book by Chapman and Cowling~\cite{cha701}:
\begin{equation}
\rho D=\frac{3}{8\sigma^2}\sqrt{\frac{k_BT}{\pi m}}\frac{1}{W_{12}^{(1)}(1)}
\end{equation}
\begin{equation}
\eta=\frac{5}{16\sigma^2}\sqrt{\frac{mk_BT}{\pi}}\frac{1}{(1/2)W_{12}^{(2)}(2)}
\end{equation}
\begin{equation}
\lambda=\frac{75}{64\sigma^2}\sqrt{\frac{k_B^3T}{\pi m}}\frac{1}{(1/2)W_{12}^{(2)}(2)}\;,
\end{equation}
where $W_{12}^{(1)}(1)$ and $W_{12}^{(2)}(2)$ are dimensionless  collision integrals~\cite{hir481} . The value of these integrals is given in~\citenum{cha701}~(page 185). 
\section{Other transport properties}
\label{sec:other_properties}

{\bf Bulk viscosity}\\Using the approach sketched in the text, we can, for instance,  obtain an expression for  the bulk viscosity, $\eta_B$. 
This quantity is not relevant in  the Chapman-Enskog approach for dilute atomic gases, because it vanishes for particles with no internal degrees of freedom. 
However, for systems with internal degrees of freedom, $\eta_B$ is finite.   
The  Green-Kubo expression for $\eta_B$ in a dilute gas is
\begin{equation}
\eta_B = \frac{1}{ 9Vk_BT}\int_0^\infty \sum_{\alpha}\langle
J_{\alpha\alpha}(0)J_{\alpha\alpha}(t) \;,
\rangle 
\end{equation}
with $\alpha$=$\{x,y,z\}$. 
The relevant flux $J_{\alpha\alpha}$ is related to the fluctuation in the trace of the stress tensor, which (for our purposes) only contains kinetic terms. 
\begin{equation}
J_{\alpha\alpha} = \sum_{i=1}^N \left(m v^{(i)}_\alpha v^{(i)}_\alpha -k_BT\right)\;.
\end{equation}
Considering, as before, first the rate of change of $J_{\alpha\alpha}$ due to individual collision events, we get (using the same notation as before:
\begin{equation}\label{eq:eta_bulk}
\dot{c}_{\eta_B} = \rho <\mu^2 v^2_{\rm rel}(0)\Delta  v^2_{\rm rel}>= -2\rho <\mu v^2_{\rm rel}(0)\Delta E^2_{\rm int}>\;,
\end{equation}
which shows that only inelastic collisions contribute to the bulk viscosity.
In what follows, we shall focus on $D$, $\eta$ and $\lambda$. 
Computing $\eta_B$ for poly-atomic molecules would require a correct description of the collision-induced transitions between different intra-molecular quantum levels. 

\noindent {\bf Thermal Diffusivity}\\The coefficient for thermal diffusion follows from the cross-correlation function of the diffusive current and the energy flux. In this case, we have to take into account fact that there are (at least) two species.
The expression for the $\dot{c}_{\rm TD}$ is
\begin{gather*}
\dot{c}_{\rm TD} =
\frac{\rho\Delta m}{3M} \left\langle  
v_{\rm rel}(0)v_{\rm rel}(1)\cos\theta\left[\frac{\Delta m}{ M}\frac{\mu}{2}v_{\rm rel}^2(1)+\frac{m_2\mathcal{E}^{(1)}(1)-m_1\mathcal{E}^{(2)}(1)}{M}\right]\right.\\
\left. -v_{\rm rel}^2(0)\left[\frac{\Delta m}{ M}\frac{\mu}{ 2}v^2_{\rm rel}+\frac{m_2\mathcal{E} ^{(1)}(1)-m_1\mathcal{E}^{(2))}(1)}{M}\right]\right\rangle
\end{gather*}
where we have used the notation $\Delta m\equiv m_2-m_1$ and $M\equiv m_2+m_1$ and $\mathcal{E}^{(i)}$ $\equiv$ $E_{\rm int}^{(i)}-h_i$, where $h_i$ is the partial molar enthalpy of species $i$.

\end{document}